\documentclass[twocolumn,showpacs,preprintnumbers,amsmath,amssymb]{revtex4}

\usepackage{graphicx}
\usepackage{dcolumn}
\usepackage{bm}

\begin{document}

\title{The hadronic SU(3) Parity Doublet Model for Dense Matter, its extension to quarks and the strange equation of state}

\author{J.~Steinheimer}
 \email{steinheimer@th.physik.uni-frankfurt.de}
\affiliation{Institut f\"ur Theoretische Physik, Goethe-Universit\"at, Max-von-Laue-Str.~1,
D-60438 Frankfurt am Main, Germany}

\author{S. Schramm}
 \email{schramm@fias.physik.uni-frankfurt.de}
\affiliation{FIAS, Johann Wolfgang Goethe University,
Frankfurt am Main, Germany}

\author{H.~St\"ocker}
\affiliation{Institut f\"ur Theoretische Physik, Goethe-Universit\"at, Max-von-Laue-Str.~1,
D-60438 Frankfurt am Main, Germany}
\affiliation{Frankfurt Institute for Advanced Studies (FIAS), Ruth-Moufang-Str.~1, D-60438 Frankfurt am Main,
Germany}
\affiliation{GSI Helmholtzzentrum f\"ur Schwerionenforschung GmbH, Planckstr.~1, D-64291 Darmstadt, Germany}

\date{\today}

\begin{abstract}
A chiral model is introduced that is based on the parity doublet
formulation of chiral symmetry including hyperonic degrees of freedom.
The phase structure of the model is determined. Depending on the masses of the chiral partners the transition
to the chirally restored phase 
shows a first-order line with critical endpoints as function of chemical potential and temperature 
in additional to the standard liquid-gas phase transition of self-bound nuclear matter.\\
We extend the parity doublet model to describe the deconfinement phase transition which is in quantitative agreement with lattice data at $\mu_B=0$. The phase diagram of the model is presented which shows a decoupling of chiral symmetry restoration and deconfinement. Loosening the constraint of strangeness conservation we also investigate the phase diagram at net strangeness density. We calculate the strangeness per baryon fraction and the baryon strangeness correlation factor, two quantities that are sensitive on deconfinement and that can be used to interpret lattice calculations.
\end{abstract}

\maketitle
\section{Introduction}
The study of dense and hot hadronic matter is a central topic of nuclear physics. 
It is directly linked to the search for the phase transition to chirally restored and deconfined matter
in ultra-relativistic heavy-ion collisions as well as to the study of extremely dense but rather cold matter inside
compact stars.
In spite of several decades of experimental and theoretical research the phase structure
of strongly interacting matter remains uncertain with the exception of the regime around cold saturated
nuclear matter and, to some extent the transition behavior at vanishing chemical potential, where lattice gauge calculation
indicate a cross-over transition to chirally restored and deconfined matter, at a temperature currently determined to be 
around 150 to 160 MeV \cite{Borsanyi:2010cj,Bazavov:2010sb}.\\
At finite chemical potential the phase structure of QCD is even less clear. While early extensions of lattice studies to finite $\mu_B$
 proposed the existence of a critical endpoint at rather small chemical potential \cite{Fodor:2001pe,Fodor:2004nz}, other lattice investigations cannot confirm evidence for its existence \cite{deForcrand:2008vr,Endrodi:2011gv}.

A central point of these investigations is the understanding of the phase transition in the hadronic and quark-hadron matter. 
Recent lattice calculations and their analysis in terms of a hadron resonance gas hint to the importance of hadronic degrees of matter in driving
the phase transition to a quark-gluon plasma \cite{Bazavov:2010sb,Huovinen:2009yb,Huovinen:2011xc}. Furthermore, the low temperature of the chiral transition \cite{Aoki:2009sc,Bazavov:2010sb}, the good agreement with chiral perturbation theory below $T_c$ \cite{Borsanyi:2010bp} and the apparent sensitivity on the hadron properties \cite{Bazavov:2010sb} (caused by lattice discretization effects) supports the idea that the chiral transition could be explained with hadronic interactions. Therefore, also a study of purely hadronic models and their properties,
especially the restoration of chiral symmetry is important.
One main benchmark for any useful comprehensive model of that kind is a reasonable description of saturated nuclear matter.
In order to have a realistic description of highly excited matter strange hadrons have to be
included in the model description.
In a simple linear sigma-model it is not possible to have
stable bound nuclear matter. Therefore a number of extended
approaches adding vector and dilaton fields were discussed \cite{Boguta:1982wr,Glendenning:1984qh,Mishustin:1993ub,Heide:1993yz}, including extensions to flavor SU(3)  \cite{Papazoglou:1997uw,Papazoglou:1998vr,Tsubakihara:2009zb}.

\section{The parity doublet model}

An elegant and alternative description of a transition to chirally restored matter is the parity doublet
model. In this approach an explicit mass term for baryons is possible, where
the signature for chiral symmetry restoration is the degeneracy of the baryons and their respective
parity partners. There are several SU(2) studies of nuclear matter adopting this approach showing that it is possible
to generate saturated matter in the parity doublet approach \cite{Detar:1988kn,Zschiesche:2006zj,Dexheimer:2007tn,Dexheimer:2008cv,Gallas:2009qp,Sasaki:2010bp}. A SU(3) parity-doublet description of hadronic matter was still missing. In \cite{Nemoto:1998um} hyperonic decays in vacuum have been studied in such an approach.
In the following we outline the basic SU(3) parity model. With this ansatz we study nuclear matter saturation in
order to fulfill one of the benchmarks for a useful model as mentioned above. Subsequently we calculate the phase diagram of isospin-symmetric
matter by varying the
baryonic chemical potential and temperature of the system. 

In the parity doublet model positive and negative parity states of the baryons
are grouped in doublets. 
The two components of the fields defining the parity partners, $\varphi_+$ and $\varphi_-$
transform in opposite way regarding chiral transformations:
\begin{eqnarray}
&\varphi'_{+R}  =  R \varphi_{+R} & \varphi'_{+L}
= L \varphi_{+L} \ \nonumber \\ 
&\varphi'_{-R} = L \varphi_{-R} & \varphi_{-L}
= R \varphi_{-L} ~,
\end{eqnarray}
where $L$ and $R$ are rotations in the left- and right handed subspaces.
This allows for a chirally invariant mass term in the
Lagrangian of the general form:
\begin{eqnarray}
&m_{0}( \bar{\varphi}_- \gamma_{5} \varphi_+ - \bar{\varphi}_+
      \gamma_{5} \varphi_- ) =  \nonumber \\
&m_0 (\bar{\varphi}_{-L} \varphi_{+R} -
        \bar{\varphi}_{-R} \varphi_{+L} - \bar{\varphi}_{+L} \varphi_{-R} +
        \bar{\varphi}_{+R} \varphi_{-L}) ,
\end{eqnarray}
where $m_0$ represents a mass parameter.
The general SU(3) extension of the approach using the non-linear representation of the fields is quite straightforward as shown in \cite{Nemoto:1998um}.
As outlined in \cite{Papazoglou:1997uw} one constructs SU(3)-invariant terms in the Lagrangian including the meson-baryon
and meson-meson self-interaction terms assuming a nonlinear realization of chiral symmetry. The part of the Lagrangian coupling the baryon and
the mesonic fields relevant in a mean-field approximation reads

\begin{eqnarray}
{\cal L_B} &=& {\rm Tr} (\bar{\Xi} i {\partial\!\!\!/} \Xi)
+ m_0 {\rm Tr}( \left(\bar{\Xi} \gamma_5 \tau_2 \Xi \right) +
D^{(1)}_s {\rm Tr}( \bar{\Xi} \left\{ \Sigma, \Xi \right\} ) \nonumber \\ &+&
F^{(1)}_s {\rm Tr}( \bar{\Xi} \left[ \Sigma, \Xi \right] )
+S^{(1)}_s {\rm Tr}( \Sigma ) {\rm Tr}( \bar{\Xi} \Xi )
 \nonumber \\ &+&
 D^{(2)}_s {\rm Tr}( \bar{\Xi} \tau_3 \left\{ \Sigma, \Xi \right\} ) +
 F^{(2)}_s {\rm Tr}( \bar{\Xi} \tau_3 \left[ \Sigma, \Xi \right] )\nonumber \\ &+&
S^{(2)}_s {\rm Tr}( \Sigma ) {\rm Tr}( \bar{\Xi} \tau_3 \Xi ) +
D_v {\rm Tr}( \bar{\Xi} \gamma_\mu \left\{ V^\mu, \Xi \right\} ) \nonumber\\
&+&
F_v {\rm Tr}( \bar{\Xi} \gamma_\mu \left[ V^\mu, \Xi \right] ) +
 S_v {\rm Tr}( V^\mu) {\rm Tr}( \bar{\Xi} \gamma_\mu \Xi ) ~.
\label{lagrangian}
\end{eqnarray}

\begin{table}[t]
\begin{center}
\caption{Model parameters for different values of the mass of the nucleonic parity partner}
\begin{tabular}{lllll}
\\
\hline\noalign{\smallskip}
$k_0$&$k_1$ &$k_2$ &$\epsilon$&\\
$(370.99 {\rm MeV})^2$ & 4.166 & -12.902& $(75.98 {\rm MeV})^4$\\
\hline \\
M$_N*$ [MeV]&1400&1440&1490&1535\\
$g_\sigma^1$&-7.27&-7.467 &-7.714&-7.933\\
$g_\sigma^8$& -0.765&-0.792&-0.823&-0.850\\
$\alpha_\sigma$&2.898&2.812&2.717&2.642\\
$g_{N\omega}$&5.375&5.375&5.5&5.563\\
\end{tabular}
\end{center}
\end{table}

\begin{figure}[b]
\centering
\includegraphics[width=0.5\textwidth]{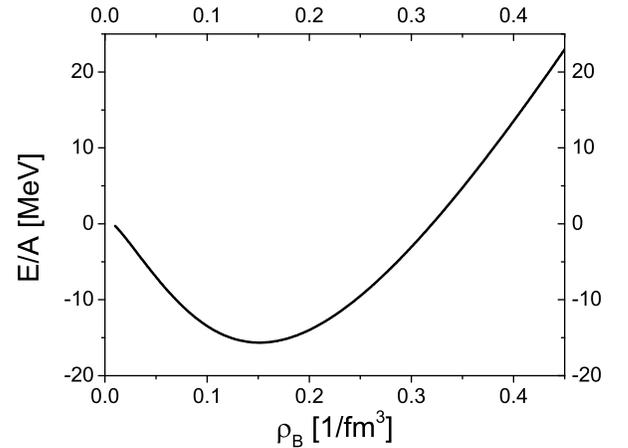}
\caption{
Negative binding energy per particle in MeV as function of density of the system for the case $M_{N*} = 1535$ MeV.  A reasonable nuclear matter ground state can be achieved.
\label{nmeos}
}
\end{figure}

Here $\Xi$ is the baryon octet whereby each field is a doublet consisting of the baryon and its negative
parity partner. $\Sigma$ and $V^\mu$ are the multiplets of the scalar and vector mesons.
The Pauli matrices $\tau_i$ act on the doublets.
In general the various sets  $D^{(i)},F^{(i)},S^{(i)}$ correspond to the D-type and F-type SU(3) invariant
baryon-meson couplings. Note that the parity doublet models allow for two different scalar coupling terms $i=1,2$. In order not to be overwhelmed by coupling constants we will restrict the set of non-zero
couplings in the actual calculations. As the term proportional to $m_0$ mixes the upper and lower components
of the parity doublets, one diagonalizes the matrix by introducing new fields $B$ with a diagonal mass matrix.
Taking along only the diagonal meson contributions, the scalar and vector condensates in the mean field approximation,
the resulting Lagrangian ${\cal L_B}$ then reads
\begin{eqnarray}
{\cal L_B} &=& \sum_i (\bar{B_i} i {\partial\!\!\!/} B_i)
+ \sum_i  \left(\bar{B_i} m^*_i B_i \right) \nonumber \\ &+&
\sum_i  \left(\bar{B_i} \gamma_\mu (g_{\omega i} \omega^\mu +
g_{\rho i} \rho^\mu + g_{\phi i} \phi^\mu) B_i \right) ~~.
\label{lagrangian2}
\end{eqnarray}
The effective masses of the baryons (assuming isospin symmetric matter) read
\begin{equation}
m^*_{i\pm} = \sqrt{ \left[ (g^{(1)}_{\sigma i} \sigma + g^{(1)}_{\zeta i}  \zeta )^2 + (m_0+n_s m_s)^2 \right]}
\pm g^{(2)}_{\sigma i} \sigma \pm g^{(1)}_{\zeta i} \zeta.
\end{equation}
where the various coupling constants $g^{(j)}_i$ are given as combinations of the original parameters $D^{(j)}, F^{(j)}, S^{(j)}$ in
equation \ref{lagrangian} and further adding a SU(3) breaking mass term that generates an explicit mass corresponding to the strangeness $n_s$ of the baryon.

\begin{figure}[t]
\begin{center}
\includegraphics[width=0.5\textwidth]{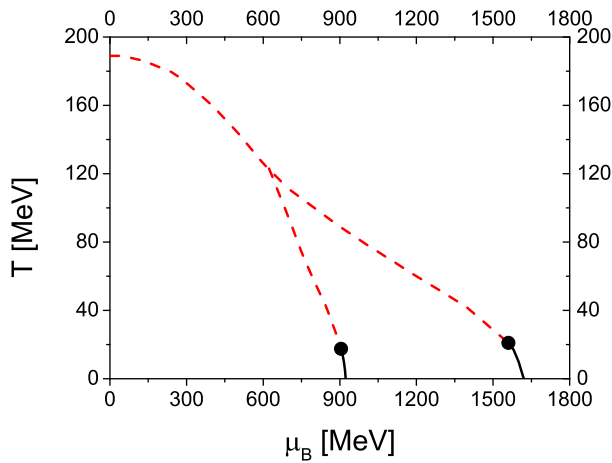}
\caption{ 
Phase transition lines of the chiral and liquid gas transitions for a mass $m_{N*} = 1535 {\rm MeV}$. Solid lines mark first-order transitions whereas dashed lines indicate a cross-over. The circles mark the critical end points. The liquid-gas and chiral symmetry restoration cross-over lines merge at higher temperature.
\label{crit}
}
\end{center}
\end{figure}

The scalar meson interaction driving the spontaneous breaking of chiral symmetry
can be written in terms of SU(3) invariants
$I_1 = Tr(\Sigma)  ~,~ I_2 = Tr(\Sigma^2) ~,~ I_3 = det(\Sigma) ~,~ I_4 = Tr (\Sigma^4) $:
\begin{equation}
V = V_0 + \frac{1}{2} k_0 I_2 - k_1 I_2^2 - k_2 I_4 + k_3 {\rm ln}(I_3)
\end{equation}
where $V_0$ is fixed by demanding a vanishing potential in the vacuum.
The explicit symmetry breaking term that generates the correct pion and kaon masses with their corresponding decay constants
can be written as
\begin{eqnarray}
&L_{SB}= m_\pi^2 f_\pi\sigma+\left(\sqrt{2}m_k^ 2f_k-\frac{1}{\sqrt{2}}m_\pi^ 2 f_\pi\right)\zeta,\nonumber&\\&
\end{eqnarray}

\begin{figure}[t]
\centering
\includegraphics[width=0.5\textwidth]{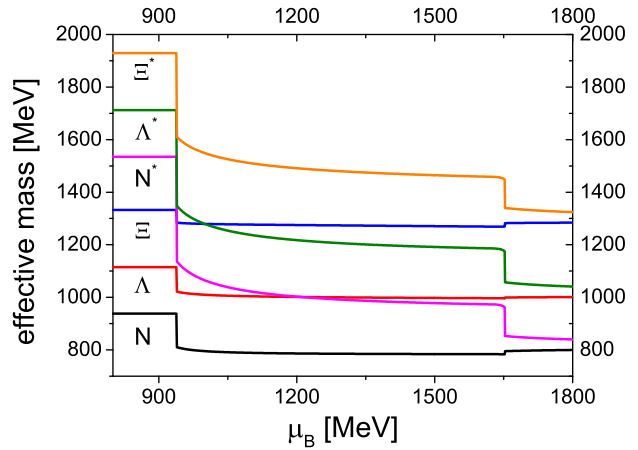}
\caption{
Masses of baryons as function of chemical potential. The degeneracy of the various parity doublets can be observed at high $\mu_B$.
\label{masses}
}
\end{figure}

The set of scalar coupling constants are fitted in order to reproduce the vacuum masses of the nucleon, and the
Lambda,  Sigma, and Xi hyperons, whereas the vector couplings are chosen to reproduce reasonable values for nuclear ground state
properties. The resulting binding energy per particle as function of density is shown in Fig. 1. For all listed parameters the ground state energy per baryon is between -15 MeV and -16 MeV, the ground state density
has a value of $\rho_0 = 0.15 fm^{-3}$ and the compressibility lies between 300 MeV and 310 MeV. The latter value is somewhat large.
Here, a more detailed and extensive parameter study might likely lead to more satisfactory values.
Note that the value of the mass parameter is set to $m_0=810$ MeV for all parametrizations. Such a choice corresponds to a rather large bare mass of the baryons. In principle such a large $m_0$ could also be generated dynamically through a coupling to the dilaton field (see e.g. \cite{Ellis:1991qx}). Such a coupling can be introduced in our model in a straight forward way. In the present investigation however we intended to have as few free parameters as possible, allthough such an extension will be subject of future investigations.

One candidate for the parity partner of the nucleon is the N(1535) resonance. However, this assignment is unclear, 
the state might also be a broad structure, so essentially
the mass of the particle (assuming its existence) is not determined. Resulting parameters for several values are shown in Table 1.
A SU(3) description, in addition to enhance the number of degrees of freedom, also necessarily increases the
number of parameters. In order not to be overwhelmed by too many new parameters, for simplicity we assume that the splitting of the
various baryon species and their respective parity partners is of the same value for all baryons, which is achieved by setting
$g^{(2)}_{\sigma i} \equiv g^{(2)}$ and $g^{(2)}_{\zeta i} = 0$. This should be sufficient for a first investigation
of the model approach. This assumption agrees quite well with the even less certain assignments of the parity partners of the hyperons.
Obvious candidates are the $\Lambda(1670)$ and $\Sigma(1750)$, whose masses roughly follow the equal splitting approximation, assuming the nucleonic
parity partner to be the  N(1535). In the case of the $\Xi^*$ the data are unclear.

\begin{figure}[t]
 \centering
\includegraphics[width=0.5\textwidth]{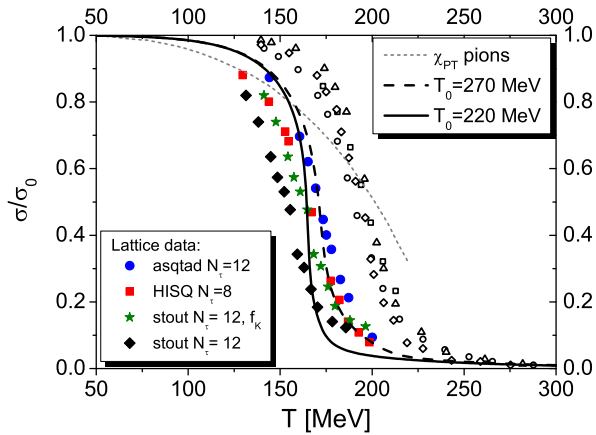}
 \caption{(Color online) Normalized value of the chiral condensate as a function of temperature, at $\mu_B=0$. The solid line is the model result for $T_0=220$ MeV and the dashed line for $T_0=270$ MeV. The grey dashed line depicts results for the pion contribution to the chiral condensate from chiral perturbation theory \cite{Gerber:1988tt}. The symbols denote lattice data, where the open symbols represent older data with the asqtad and p4 action and the colored symbols are more recent results (see text).}
 \label{1}
\end{figure}

In another simplification the hyperonic vector interactions were tuned to generated reasonable
optical potentials of the hyperons in ground state nuclear matter, with $U_\Lambda(\rho_o) ~ -28\,{\rm MeV}$ and $U_\Xi(\rho_o) ~  -18\,{\rm MeV}$ . The value for the strange quark mass was fixed at $m_s = 150$\,MeV. The numbers used are summarized in Table 1.
A more exhaustive study of various parameter setups will be performed in future work.

The equations of motion following from Eqs. (4, 6, 7)
are then solved self-consistently in mean field approximation
by minimizing the grand canonical potential as function of baryonic chemical
potential and temperature.

The resulting phase diagram for the transition from chirally broken to chirally restored phase is shown
in Fig. 2. One can observe two first-order transition lines at high densities. The first one is the
liquid-gas phase transition, which indicates that the model exhibits a bound nuclear ground state. The second
one is the chiral transition, also signaling the onset of the population of the baryonic parity partners. The lines stop at second-order
critical end points, at (T$_c$,$\mu_c$) = (21 MeV, 1560 MeV) and (17.5 MeV, 905 MeV). The critical point of the chiral transition is very low in temperature.
 At values of  $M_{N*}$ below 1460 MeV the first-order transition becomes a cross-over for all values of T and $\mu$.
 Both cross-over lines join at a temperature $T \approx 120$\,MeV.
 
Fig. 3 shows the effective masses of the baryons with their parity partners as function of baryochemical potential.
One can observe the effect of the two phase transitions, leading to essentially degenerate opposite parity states.

Here the calculations were done for isospin symmetric matter with strangeness zero, which would be the logical
first-order assumption for matter created in heavy-ion collisions.
The study of star matter in beta equilibrium including leptons for ensuring charge neutrality is in progress. Results will be
presented in a forthcoming publication.

\section{Including Mesons and Quarks}

\begin{figure}[t]
 \centering
\includegraphics[width=0.5\textwidth]{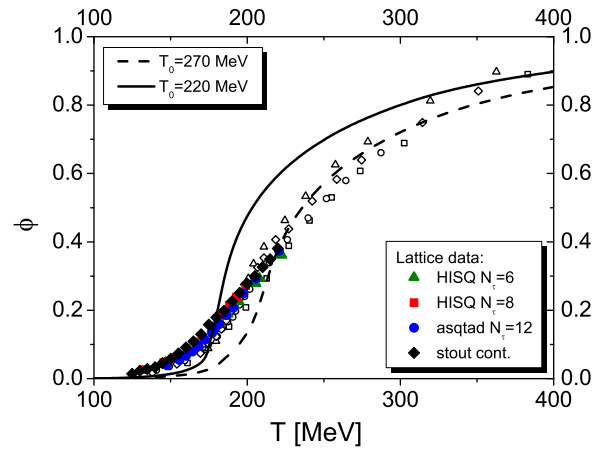}
 \caption{(Color online) Value of the Polyakov loop as a function of temperature at $\mu_B=0$. The solid line is the model result for $T_0=220$ MeV and the dashed line for $T_0=270$ MeV. The symbols denote lattice data, where the open symbols represent older data with the asqtad and p4 action and the colored symbols are more recent results.}
 \label{2}
\end{figure}

Recent results from different lattice collaborations indicate that the chiral phase transition at $\mu_q=0$ occurs at a rather low temperature. 
Although there are still considerable systematic lattice effects there seems to be a consensus that the crossover temperature is between $150$ and $160 $ MeV. 
This can be seen in figure (\ref{1}) where the symbols denote the different lattice actions. In particular intriguing is the fact that the value of the chiral
condensate already drops to about $80$ to $90 \%$ of its vaccum value at temperatures less than $150$ MeV. In the parity doublet model such an early 
decrease of the chiral condensate can be hardly accommodated by only coupling the baryons to the chiral fields as their mass is rather
large and they are not thermodynamically activated at such low temperatures. On the other hand calculations in chiral perturbation theory 
have shown that the pionic contribution to the chiral condensate is considerable at such low temperatures because of the small pion mass. 
This is also shown in figure (\ref{1}), as the grey dashed line depicts the chiral perturbation theory results for the pion self interaction taken from \cite{Gerber:1988tt}. 
One can see, that at low temperature the behavior of the chiral condensate seems dominated by the pseudoscalar contributions while only at larger 
temperature the baryon interactions become important. Consequently we would like to include effects of the pseudoscalars in our parity
doublet model to accommodate the low temperature behavior of the chiral condensate at small net baryon densities. To this end we take into account the 
coupling of the scalar field, which originates from the explicit symmetry breaking term, Eq. (7) (see Ref. \cite{Mishra:2008kg}). Including the pseudoscalar 
mesons this term generates a mass for the pions as $m_\pi^2 = m_{\pi,0}^2 \sigma / \sigma_0$. This leads to an increase of the pressure of the pion gas with 
decreasing scalar condensate, thus driving the phase transition to lower temperatures. 

\begin{figure}[t]
 \centering
\includegraphics[width=0.5\textwidth]{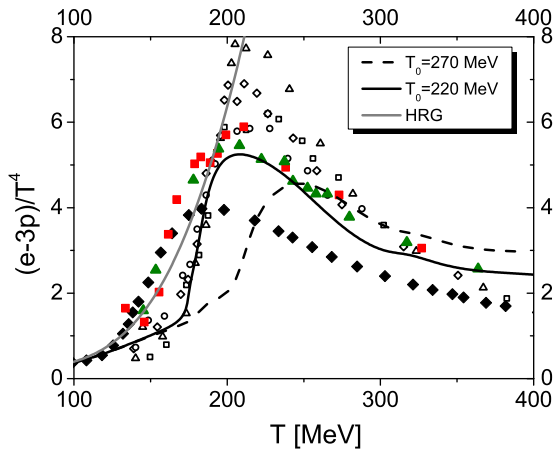}
 \caption{(Color online) The interaction measure (defined as $(e-3p)/T^4$) as a function of temperature at $\mu_B=0$. The solid line is the model result for $T_0=220$ MeV and the dashed line for $T_0=270$ MeV. The symbols denote lattice data, where the open symbols represent older data with the asqtad and p4 action and the colored symbols are more recent results. The grey line represents the interaction measure for a hadron resonance gas equation of state which includes all hadronic resonances up to 2.2 GeV.}
 \label{3}
\end{figure}

It is well known that at some temperature QCD exhibits a transition to a deconfined phase at which the quarks become the dominant degrees of freedom. When this deconfinement will appear and what the order parameter for this transition might be is still under heavy debate \cite{Aoki:2006we,Bazavov:2010bx}. Assuredly one can only say that it occurs in a temperature region of $T_{dec}\approx 160 - 400$ MeV. Nevertheless at some point the hadronic parity doublet model will not be the appropriate effective description of QCD and one needs to introduce a deconfinement mechanism in the model. In this work we will apply a mechanism that has been introduced in \cite{Steinheimer:2010ib} to add a deconfinement transition in a chiral hadronic model. This is done by adding an effective quark and gluon contribution as done in the PNJL approach \cite{Fukushima:2003fw,Ratti:2005jh}. This model uses the Polyakov loop $\Phi$ as the order parameter for deconfinement. $\Phi$ is defined via $\Phi=\frac13$Tr$[\exp{(i\int d\tau A_4)}]$, where $A_4=iA_0$ is the temporal component
of the SU(3) gauge field, distinguishing $\Phi$, and its conjugate $\Phi^{*}$ at finite baryon densities \cite{Fukushima:2006uv,Allton:2002zi,Dumitru:2005ng}. In recent years the PNJL model has been widely used and extended to include non-local interactions as well as an imaginary chemical potential (see also \cite{Ratti:2004ra,Roessner:2006xn,Sasaki:2006ww,Ratti:2007jf,Rossner:2007ik,Ciminale:2007sr,Schaefer:2007pw,Fu:2007xc,Hell:2008cc,Abuki:2008nm,Fukushima:2008wg,Fukushima:2008is,Costa:2008gr,Costa:2008dp,Hansen:2006ee,Mukherjee:2006hq,Abuki:2008tx,Abuki:2008ht,Fukushima:2009dx,Mao:2009aq,Schaefer:2009ui,Hell:2009by,Contrera:2009hk,Radzhabov:2010dd,Contrera:2010kz,Herbst:2010rf,Pagura:2011rt,Kashiwa:2011td,Weise:2010cp,Blaschke:2010zz}).\\
The effective masses of the quarks
are generated by the scalar mesons except for a small explicit
mass term ($\delta m_q=5$ MeV and $\delta m_s=150$ MeV for the strange quark) and $m_0$ (explain):
\begin{eqnarray}
&m_{q}^*=g_{q\sigma}\sigma+\delta m_q + m_{0q},&\nonumber\\
&m_{s}^*=g_{s\zeta}\zeta+\delta m_s + m_{0q},&
\end{eqnarray}
with values of $g_{q\sigma}=g_{s\zeta}= 4.0$. As in the case of the baryons we also introduced a mass parameter $m_{0q}= 200$ MeV for the quarks. Again this additional mass term can be due to a coupling of the quarks to the dilaton field (gluon condensate). Fot this mass term the quarks do not appear in the nuclear ground state which would be an unphysical result. This allows to set the vector type repulsive interaction strength of the quarks to zero. A non-zero vector interaction strength would lead to a massive deviation of the quark number susceptibilities to lattice data as has been indicated in different mean field studies \cite{Kunihiro:1991qu,Ferroni:2010xf,Steinheimer:2010sp}.\\
A coupling of the quarks to the Polyakov loop is introduced in the thermal energy of the quarks. Their thermal contribution to the grand canonical potential $\Omega$, can be written as:
\begin{equation}
	\Omega_{q}=-T \sum_{i\in Q}{\frac{\gamma_i}{(2 \pi)^3}\int{d^3k \ln\left(1+\Phi \exp{\frac{E_i^*-\mu_i}{T}}\right)}}
\end{equation}
and
\begin{equation}
	\Omega_{\overline{q}}=-T \sum_{i\in Q}{\frac{\gamma_i}{(2 \pi)^3}\int{d^3k \ln\left(1+\Phi^* \exp{\frac{E_i^*+\mu_i}{T}}\right)}}
\end{equation}

\begin{figure}[t]
 \centering
\includegraphics[width=0.5\textwidth]{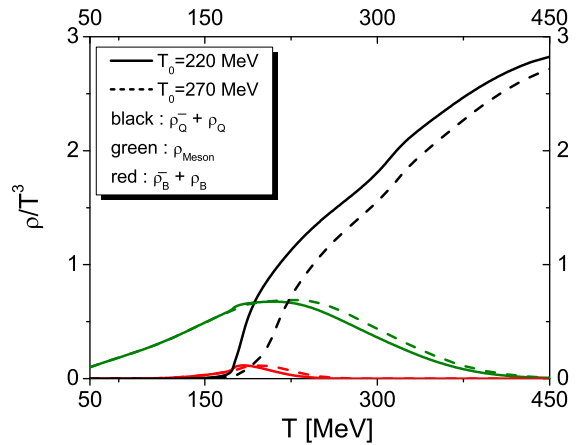}
 \caption{(Color online) Number densities of the different particle species as a function of temperature at $\mu_B=0$. The solid lines show results for $T_0=220$ MeV and the dashed line for $T_0=270$ MeV. The mesons are shown in green, the densities for baryons plus anti-baryons in red and the quarks plus anti-quarks in black.}
 \label{4}
\end{figure}

The sums run over all quark flavors, where $\gamma_i$ is the corresponding degeneracy factor, $E_i^*=\sqrt{m_{i}^{*2}+p^2}$ the energy and $\mu_i^*$ the chemical potential of the quark.\\
All thermodynamical quantities, energy density $e$, entropy density $s$ as well as the
densities of the different particle species $\rho_i$, are derived from the grand canonical potential.
It includes the effective potential $U(\Phi,\Phi^*,T)$, which controls the dynamics of the Polyakov-loop.
In our approach we adopt the ansatz proposed in \cite{Ratti:2005jh}:
\begin{eqnarray}
	U&=&-\frac12 a(T)\Phi\Phi^*\nonumber\\
	&+&b(T)ln[1-6\Phi\Phi^*+4(\Phi^3\Phi^{*3})-3(\Phi\Phi^*)^2]
\end{eqnarray}
 with $a(T)=a_0 T^4+a_1 T_0 T^3+a_2 T_0^2 T^2$, $b(T)=b_3 T_0^3 T$.\\
The parameters $a_0, a_1, a_2$ and $b_3$ are fixed, as in \cite{Ratti:2005jh}, by demanding a first order phase transition in the pure gauge sector at $T_0=270$ MeV, and that the Stefan-Boltzmann limit of a gas of gluons is reached for $T \rightarrow \infty$. In general of course the presence of quarks may have a significant influence on the Polyakov potential \cite{Schaefer:2007pw} so one should not regard the parameters to be absolutely fixed.\\

In the following we introduce excluded volumes for the hadrons in the system. As a onsequence the hadronic contributions from the equation of state at high temperatures and densities will be suppressed.
Including effects of finite-volume particles in a thermodynamic model for hadronic matter, was proposed long ago \cite{Hagedorn:1980kb,Baacke:1976jv,Gorenstein:1981fa,Hagedorn:1982qh,Rischke:1991ke,Cleymans:1992jz,Kapusta:1982qd,Bugaev:2000wz,Bugaev:2008zz,Satarov:2009zx}. In recent publications \cite{Steinheimer:2010ib,Steinheimer:2010sp} we adopted this ansatz to successfully describe a smooth transition from a hadronic to a quark dominated system (see also \cite{Sakai:2011fa}).\\

In particular we introduce the quantity $v_{i}$ which is the volume excluded of a particle of species $i$ where we only distinguish between hadronic baryons, mesons and quarks. Consequently  $v_{i}$ can assume three values:
\begin{eqnarray}
 v_{Quark}&=&0 \nonumber \\
 v_{Baryon}&=&v \nonumber \\
 v_{Meson}&=&v/a \nonumber
\end{eqnarray}

Where $a$ is a number larger than one. In our calculations we choose a value of $a=8$, which assumes that the radius $r$ of a meson is half of the radius of a baryon.
Note that at this point we neglect any possible density-dependent and Lorentz contraction effects on the excluded volumes as introduced in \cite{Bugaev:2000wz,Bugaev:2008zz}.

The modified chemical potential $\widetilde{\mu}_i$, which is connected to the real chemical potential $\mu_i$ of the $i$-th particle species, is obtained by the following relation:
\begin{equation}
	\widetilde{\mu}_i=\mu_i-v_{i} \ P
\end{equation}
where $P$ is the sum over all partial pressures. To be thermodynamically consistent, all densities ($\widetilde{e_i}$, $\widetilde{\rho_i}$ and $\widetilde{s_i}$) have to be multiplied by a volume correction factor $f$, which is the ratio of the total volume $V$ and the reduced volume $V'$, not being occupied:
\begin{equation}
	f=\frac{V'}{V}=(1+\sum_{i}v_{i}\rho_i)^{-1}
\end{equation}

\begin{equation}
e=\sum_i f \ \widetilde{e_i} \quad
\rho_i=f \ \widetilde{\rho_i} \quad
s=\sum_i f \ \widetilde{s_i}
\end{equation}

As a consequence, the chemical potentials of the hadrons are decreased by the quarks, but not vice versa. In other words as the quarks start appearing they effectively suppress the hadrons by changing their chemical potential, while the quarks are only affected through the volume correction factor $f$.\\

\begin{figure}[t]
 \centering
\includegraphics[width=0.5\textwidth]{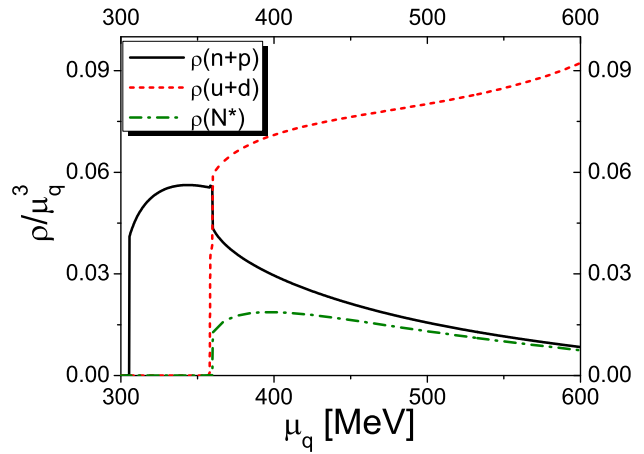}
 \caption{(Color online) Densities of the different particle species at $T=0$ and as a function of $\mu_q=\mu_B/3$. The black solid line depicts the density of protons plus neutrons, the red dashed line for up plus down quarks and the green dash dotted line for the chiral partners of the nucleons.}
 \label{6}
\end{figure}

\begin{figure*}[t]
 \centering
\includegraphics[width=\textwidth]{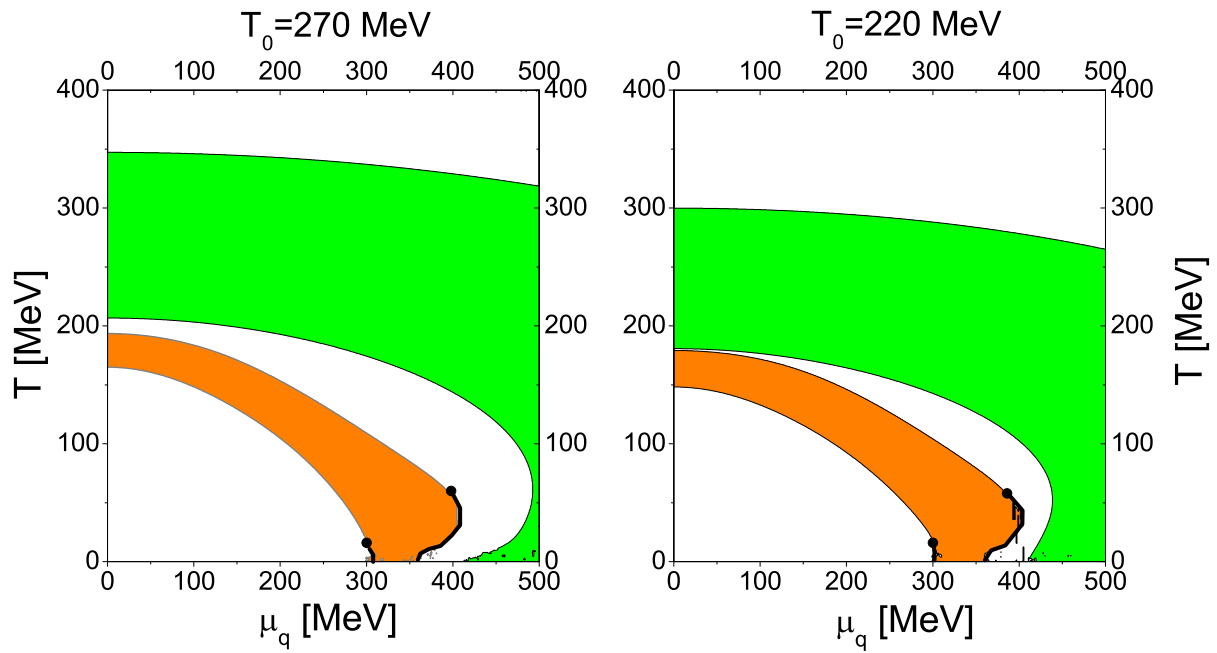}
 \caption{(Color online) Phase diagrams for our model with two different values of $T_0$. Within the orange region the normalized value of the chiral condensate lies between $0.2<\sigma/\sigma_0<0.8$. In the green area the value of the Polyakov loop is in between $0.2<\phi<0.7$. The black lines indicate first order phase transitions, where the points indicate the critical endpoints. }
 \label{7}
\end{figure*}

\section{Thermodynamic Properties}\label{results}

A surprising result from recent lattice studies of the 2+1 flavor QCD equation of state at finite temperature is the apparent decoupling of the chiral phase transition and the increase of the Polyakov loop which was long thought to be a good order parameter for deconfinement. In particular one observes that the steepest change in the chiral condensate occurs at a low temperature of roughly $150-160 $ MeV, depending on the choice of the lattice action as well as the scale which is used to translate lattice quantities into the physical temperature. On the other hand a considerable increase in the value of the Polyakov Loop is only observed above $T\approx 210$ MeV, where this result seems quite independent of the lattice action that is applied.\\
In the following we will compare results on the order parameters and thermodynamic quantities calculated within our model to recent lattice results from the different collaborations \cite{Cheng:2009zi,Aoki:2009sc,Bazavov:2009zn,Borsanyi:2010cj,Bazavov:2010sb,Bazavov:2010pg,Petreczky:2009at,Kaczmarek:2002mc,Borsanyi:2010bp}
Figure \ref{1} displays the results for the temperature dependence of the expectation value of the $\sigma$ field, normalized to its ground state value, from the SU(3) parity model including quarks. The dashed line depicts the results when the parameter of the Polyakov potential is unchanged ($T_0 = 270 $ MeV) and the solid line when $T_0$ is changed to $220$ MeV. Note that such readjustments of $T_0$ are commonly used in a number of PNJL studies and $T_0$ can even depend on $\mu_B$ \cite{Herbst:2010rf,Fukushima:2010pp}.

Our results are compared to calculations from chiral perturbation theory (grey short dashed line) \cite{Gerber:1988tt} and recent lattice results (colored symbols depict the more recent results while the open symbols refer to previously used lattice actions).\\
At temperatures below $160$ MeV the decrease of the $\sigma$ is dominantly caused by the pseudoscalar-scalar coupling. Compared to the chiral perturbation theory results, which depict the pion contribution to the chiral condensate, our model still shows a slower decrease of the chiral condensate with temperature, while lattice data are rather well described by the chiral perturbation theory up to temperatures of 150 MeV. Above that temperature the baryon scalar interaction (and at even higher temperature the quark scalar interaction) starts to contribute to the value of the $\sigma$-field. In this regime our model gives a good description of the lattice data. The chiral critical temperature $T_c^{\chi}$ only weakly depends on the value of $T_0$ because chiral symmetry restoration is mainly driven by the hadronic interactions with the fields. We obtain the values of $T_c^{\chi}$ from the maximum in $\partial \sigma/ \partial T$ as $T_c^{\chi}= 172$ and $165$ MeV  respectively.\\

While the temperature dependence of the lattice results on the chiral condensate strongly depends on the action and lattice spacing that is applied, this dependence is not so strong for the Polyakov loop as can be seen in figure \ref{2}. Here again we compare our model results with lattice data. Because $T_0$ directly influences the dynamics of the Polyakov loop we see a strong dependence of $T_c^{PL}$ on the change in $T_0$. We obtain $T_c^{PL}= 210$ and $173$ MeV.
the general shape of the curve of $\phi(T)$ is not changed considerably with $T_0$ and we observe that we cannot accommodate the almost linear increase of the Polyakov loop at lower temperatures, as seen on the lattice, by simply adjusting $T_0$.\\

Recent lattice studies start to agree on the temperature dependence of the order parameters. However this is not the case when the interaction measures, closely related to the thermodynamic properties of the matter, are compared. In figure \ref{3} we compare our model results on the interaction measure, for the two values of $T_0$, with different lattice data sets. First we have to note that the results from the HotQCD collaboration (red squares and green triangles) differ considerably from those of the Wuppertal-Budapest group (black diamonds). Furthermore we see that the interaction measure for our calculation with $T_0=220$ MeV gives a drastically better agreement with the HotQCD lattice results than for $270$ MeV. in any case our model underestimates the contribution to the interaction measure at low temperatures. However, this could be understood by missing contributions from hadronic resonances as is seen from the grey line which depicts $(e-3 p)/T^4$ for a hadron resonance gas which has already been shown to give good results for the interaction measure at temperatures below $T_c^{\chi}$.\\
The transition from a hadron to quark dominated system is depicted in figure \ref{4} with the different particle number densities at $\mu_B = 0$. While the mesons dominate the low temperature region (green lines) one can clearly see that the hadrons are slowly removed from the system as the quark density increases (black lines) and consequently the hadrons play less of a role for the thermodynamic quantities.\\

\subsection{Finite $\mu_B$} 

\begin{figure}[t]
 \centering
\includegraphics[width=0.5\textwidth]{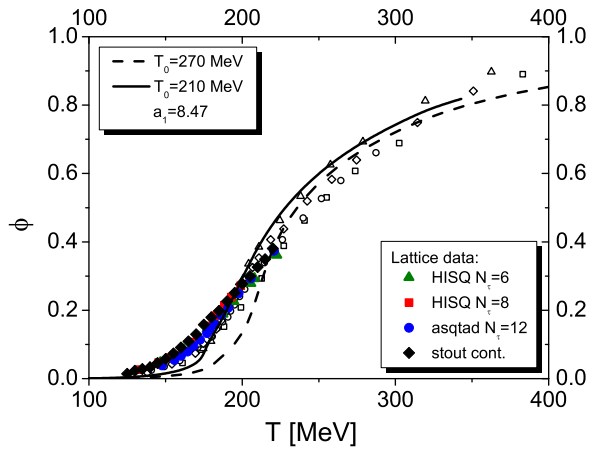}
 \caption{(Color online) Value of the Polyakov loop as a function of temperature at $\mu_B=0$. This time the solid line is the model result for $T_0=210$ MeV and a modifies parameter $a_1=-8.47$. The dashed line stands for the previous result with $T_0=270$ MeV and $a_1=-2.47$. The symbols denote lattice data, where the open symbols represent older data with the asqtad and p4 action and the colored symbols are more recent results.}
 \label{9}
\end{figure}

\begin{figure}[t]
 \centering
\includegraphics[width=0.5\textwidth]{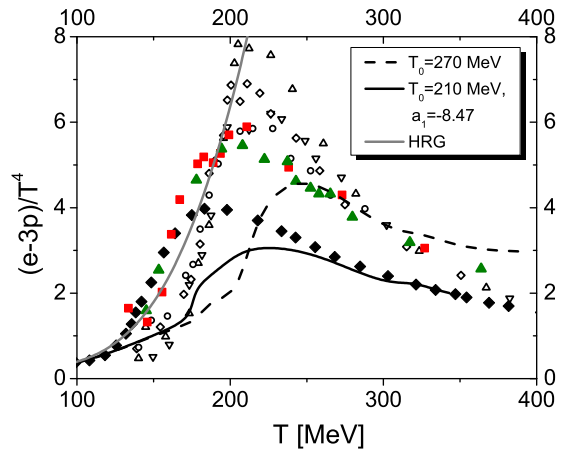}
 \caption{(Color online) The interaction measure (defined as $(e-3p)/T^4$) as a function of temperature at $\mu_B=0$. This time the solid line is the model result for $T_0=210$ MeV and a modifies parameter $a_1=-8.47$. The dashed line stands for the previous result with $T_0=270$ MeV and $a_1=-2.47$. The symbols denote lattice data, where the open symbols represent older data with the asqtad and p4 action and the colored symbols are more recent results. The grey line represents the interaction measure for a hadron resonance gas equation of state which includes all hadronic resonances up to 2.2 GeV. }
 \label{10}
\end{figure}

An advantage of our effective model is that, unlike lattice studies, we can easily extend our studies to finite baryon densities and explore the phase behavior at $\mu_B> 0$. Figure \ref{6} shows again the number densities of different hadrons and quarks, this time  at $T=0$, and as a function of $\mu_B$. at the liquid-gas phase transition the nucleon density exhibits a jump as expected. At even higher chemical potentials we can observe a second jump in the densities. at this point the chiral partner of the nucleon is activated, as well as the quark degrees of freedom. Both steps in the densities correspond to jumps in the chiral condensate as can be seen more clearly in figure \ref{7}. In this figure we depict the phase diagrams obtained from our model for the two values of $T_0$. The black lines with endpoints depict the region where the change of the chiral condensate is of first order. Instead of drawing an ambiguous crossover 'line' we show the regions in which the value of the chiral condensate changes from $0.2<\sigma/\sigma_0<0.8$ as the orange area. This region we will refer to as the chiral crossover region. The green area is defined as the region where the Polyakov loop is between $0.2<\phi<0.8$, which we will refer to as the deconfinement crossover. Please note that within our model the Polyakov loop generally changes in a smooth crossover and only exhibits a first order phase transition at the chiral phase transition, where it jumps from $\phi=0$ to $0.1$. The position of the chiral critical endpoint is $T_{cep}=58$ MeV and $\mu_B^{cep}\approx 1200$ MeV for $T_0=270$ MeV and for $T_0=220$ MeV only the chemical potential changes to $\mu_B^{cep}\approx 1150$ MeV. Such a low temperature endpoint as well as the fact that the crossover region does not become considerably wider with increasing chemical potential both agree well with recent lattice findings \cite{Endrodi:2011gv}. Generally the deconfinement crossover occurs at a higher temperature (higher chemical potential) as the chiral phase transition. This gives rise to an interesting phase of chirally symmetric confined hadronic matter.

\begin{figure*}[t]
 \centering
\includegraphics[width=\textwidth]{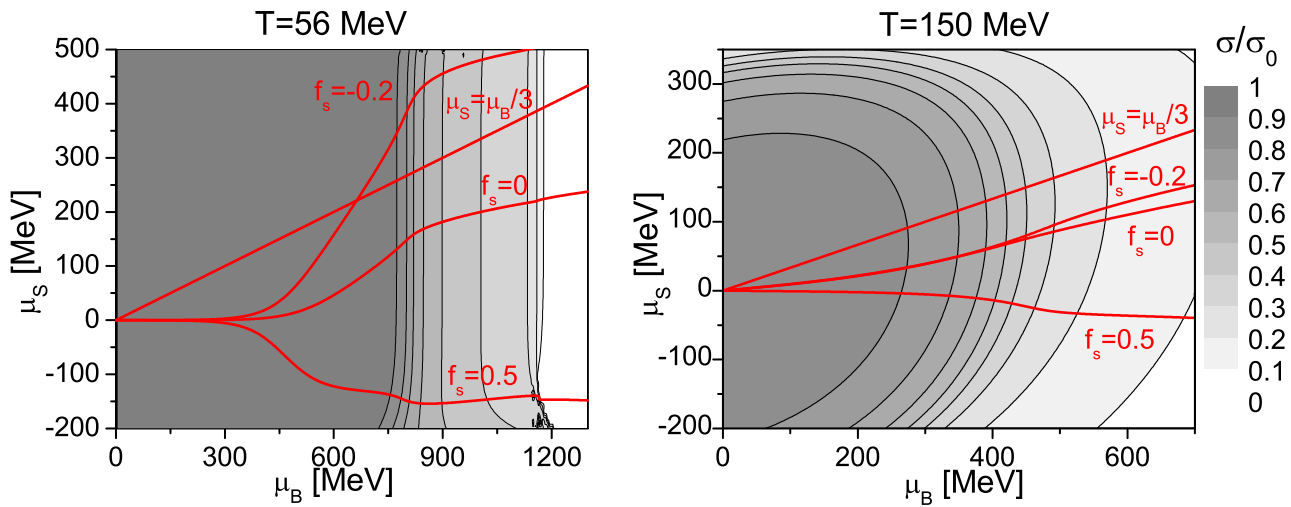}
 \caption{(Color online) Contour plots of the normalized chiral condensates as a function of the chemical potentials $\mu_B$ and $\mu_S$ for fixed temperature. The red lines correspond to different values of a fixed strangeness to baryon fraction $f_s$.}
 \label{11}
\end{figure*}

\subsection{Varying Polyakov Loop parameters}

As had been mention in the previous section, our model does not yield a good description of the temperature behavior of the Polyakov loop together with a good description of the interaction measure. Within the PNJL model it has also been pointed out that a concurrent description of both order parameters and the interaction measure is usually not achieved when simply $T_0$ is varied (see e.g. \cite{Hell:2011ic}).\\
In our model in particular the linear increase of the Polyakov loop is ill described. Usually the parameters of the Polyakov potential are fitted to pure gauge lattice result. However it has been shown that the presence of quarks may have an influence on these parameters \cite{Schaefer:2007pw} and in general there is no reason that only $T_0$ should be affected by such a quark coupling. To investigate and illustrate what effects a different parameter set of the Polyakov potential has on the interaction measure we will adjust one parameter ($a_1$) to $a_1=-8.47$ (instead of $a_1=-2.47$). Together with changing $T_0$ to $T_0=210$ MeV this gives an improved description of the lattice data for the Polyakov loop behavior as can bee seen in figure \ref{9}.\\
Figure \ref{10} shows the resulting interaction measure for our adjusted parameter. One can see that, due to the slower rise of the Polyakov loop, the peak in the interaction measure is considerably lowered. Taking into account missing contributions from resonances one can even conclude that this parametrization compares more favorably with the Wuppertal-Budapest results than with the HotQCD lattice data.\\
Consequently, to understand the interplay between the order parameters and the interaction measure from lattice calculations, one has to resolve the still existing discrepancies. Only then conclusions regarding the role of the degrees of freedom in the chiral and deconfinement phase transition can be drawn.

\section{The strange Phase diagram}

Until now all calculations where restricted to the limit of vanishing net strange density. Because the strong interaction conserves the net strange particle number, the equation of state used for the description of heavy ion collisions is usually considered to be net strange free. However, there are several issues that make the study of the strange EoS interesting. Some of these issues are:

\begin{enumerate}
\item As has been shown in \cite{Steinheimer:2008hr} the net strangeness distribution in coordinate and momentum space of a heavy ion collision can fluctuate, although the total net strangeness is zero. To dynamically treat such a system, and calculate observables that arise from such a strangeness fluctuation, the equation of state for $\rho_{s}\ne 0$ needs to be evaluated.

\item Compact stars are very dense and long lived objects. Due to a $\beta$-equilibrium inside the star, net-strange conservation can be violated by the weak interaction.

\item Lattice QCD results at finite $\mu_B$ are often evaluated through a Taylor expansion in $\mu_B$ at $\mu_B=\mu_S=0$. A vanishing strange number chemical potential usually induces a non-vanishing net strangeness, which means that the equation of state of net-strange matter is calculated.
\end{enumerate}

First investigations on the strange equation of state were done in \cite{Lee:1992hn}, where one usually considered a first order transition from a hadron to a quark phase. In our model we are able to discuss the strange EoS in the context of a smooth transition from a confined hadron phase to a deconfined quark phase.\\
Figure \ref{11} presents our results on the order parameter of the chiral phase transition as a function of $\mu_B$ and $\mu_S$ at fixed temperature. The red lines indicate paths of constant values for $f_s=\rho_s/\rho_B$, the strangeness per baryon fraction. note that $f_s=0$ corresponds to our results in section \ref{results} (with $T_0= 220$ MeV). At the temperature $T=56$ MeV, the critical endpoint of the chiral phase transition was located at $\mu_B^{cep}\approx 1150$ MeV. We can observe that for increasing $f_s$, the change in the order parameter becomes steeper and the value of $T_{CEP}$ increases slightly to $T_{CEP}= 68$ MeV for $f_s=0.5$. 
At the larger temperature we also observe a slight change in the phase structure. Here, for increasing $f_s$, the crossover moves closer to the $\mu_{B,S}=0$-line.\\

For a gas of deconfined quarks there is a strong correlation between the baryon number and strangeness. In a hadronic medium such a correlation is usually not trivial as strangeness can be found in mesons and baryons. These considerations led to the idea that the so called strangeness-baryon correlation factor $c_{BS}$ is sensitive to the deconfinement and/or chiral phase transition \cite{Koch:2005vg}. On the other hand the strangeness to baryon ratio $f_s$ should also be sensitive on any phase transition at finite baryon densities.\\
On the lattice such quantities are usually calculated as functions of the expansion coefficients, it is defined as \cite{Koch:2005vg}:
\begin{equation}
	c_{BS}=-3\frac{\left\langle N_B N_S\right\rangle- \left\langle N_B \right\rangle \left\langle N_S\right\rangle}{\left\langle N_S^2 \right\rangle - \left\langle N_S \right\rangle^2}
\end{equation}

\begin{figure}[t]
 \centering
\includegraphics[width=0.5\textwidth]{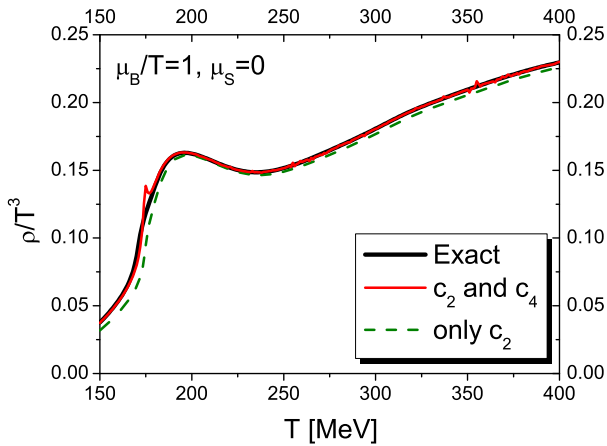}
 \caption{(Color online) Baryon number density divided by $T^3$ as a function of temperature for $\mu_B/T=1$. Shown is the exact solution from the model (black solid line) and Taylor expansions of the density, taking into account only the second (green dashed line) and forth (red solid line) coefficient.}
 \label{12}
\end{figure}

\begin{figure}[t]
 \centering
\includegraphics[width=0.5\textwidth]{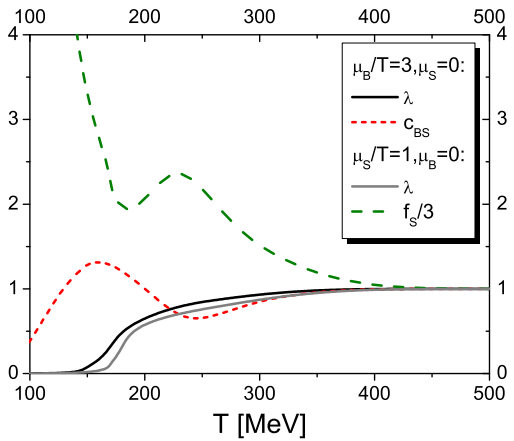}
 \caption{(Color online) Shown are the strangeness to baryon correlation coefficient $c_{BS}$ (red short dashed line) compared to the quark-gluon fraction $\lambda=e_{Quarks+Gluons}/e_{Tot}$ (black solid line) as a function of temperature for $\mu_B/T=3$ and $\mu_S=0$. The plot also shows the strangeness per baryon fraction $f_s$ (green dashed line) and the quark-gluon fraction $\lambda$ (grey solid line) as a function of temperature for $\mu_S/T=1$ and $\mu_B=0$.}
 \label{13}
\end{figure}

The question is how many coefficients are needed to evaluate the baryon and/or strange densities at finite $\mu_B/T$ to a given accuracy \cite{Karsch:2010hm}. As an example figure \ref{13} shows the baryon density as a function of temperature at fixed $\mu_B/T=1$ and $\mu_S=0$ for our exact model calculation. Alternatively we can also numerically extract the expansion coefficients for our model and expand the density in powers of $\mu/T$. One can see that already the result, taking into account only the first non vanishing coefficient $c_2$, gives a quite reasonable description of the exact result. Taking into account the 2nd and 4th order coefficient already allows to describe the exact result to high accuracy, except at the point of the crossover transition. This means that, in order to calculate $f_s$ at finite $\mu_S / T$, it is sufficient to extract the coefficients up to 4th order ($c^{B,S}_{4,4}$) from the lattice.\\

The information that can be extracted from these quantities is exemplified in figure \ref{13}. Here we show the exact solution for $c_{BS}$ as a function of temperature for $\mu_B/T=3$ and $\mu_S=0$. One can observe a distinct peak at $T\approx 150$ MeV $\Rightarrow \mu_B=450$ MeV. Comparing with figures \ref{7} and \ref{11} one can identify this peak with the crossover transition of the chiral condensate. Such a behavior of $c_{BS}$ has been predicted and also has been shown to exist in lattice data \cite{Schmidt:2009qq}. At higher temperatures the strangeness to baryon correlation approaches unity which resembles closely the behavior of the quark and gluon fraction $\lambda=e_{Quarks+Gluons}/e_{Tot}$ of the system. In comparison figure \ref{13} also shows the temperature dependence of $f_s$ at $\mu_S/T=1$ and $\mu_B=0$. This quantity is even more sensitive in the quark-gluon fraction as is $c_{BS}$, while it seems to be not very sensitive to the chiral phase transition. The peak in $f_s$ can rather be understood as a consequence of our excluded volume treatment, where mesons have a smaller
excluded volume than baryons. Hence mesons, that can carry strangeness, are less suppressed than baryons.

\section{Discussion}
We presented results on the phase structure of a SU(3) parity-doublet description of hot and dense hadronic matter.
With appropriate parameters we could generate a quantitatively acceptable nuclear ground state. The phase diagram in 
temperature and baryochemical potential exhibits a liquid-gas first-order phase transition as well as a chiral phase transition 
that is connected to the population of the parity partners  and the onset of their degeneracy with the normal baryon states. Depending on the mass gap between the baryons and their
parity partners this transition is first-order at high densities and low temperatures and a crossover otherwise, or a smooth crossover for all values
of $T, \mu$ for smaller mass gaps. In order not to be overwhelmed by too many new parameters some simplifications of the parameter choice have been made, assuming an equal mass gap between all positive and negative parity baryons. These restrictions should be relaxed in further studies to explore 
the model in more detail. 

In the second part of this paper we extended the SU(3) parity doublet model to incorporate a deconfinement phase transition. When comparing our results to lattice data at $\mu_B=0$ we find that the low temperature behavior of the chiral condensate is dominated by hadronic interactions. In such a scenario a decoupling of the Polyakov loop and the chiral condensate, as is seen in recent lattice studies, can be easily understood. A feature which is common to PNJL-type models is that a simultaneous description of the interaction measure and the Polyakov loop cannot be achieved simply by adjusting the parameter $T_0$. If we loosen also the constraint on other parameters of the Polyakov loop potential we obtain an improved description of the Polyakov loop dynamics. It would be interesting to investigate if, e.g. in PNJL models, the slower increase also shows to have drastic effects on the interaction measure as has the parameter change presented in this work. As lattice results still differ strongly in their results on the interaction measure it is not possible to say if such a reparametrization improves or weakens the model. Consequently it is of utmost importance to understand and settle the differences in the lattice results to be able to understand the interplay between the order parameters and the thermodynamics, i.e. the active degrees of freedom.\\

At finite baryon densities our model describes the deconfinement transition as a continuous crossover for all values of $\mu_B$. Only the chiral order parameter exhibits two discontinuities. One is related to the nuclear liquid gas phase transition while the other can be identified as the chiral phase transition and appears at larger densities. We also observe that the critical endpoint of the chiral phase transition has a rather small temperature $T_{cep}= 56$ MeV. As the chiral phase transition is driven mainly by hadronic interactions and the deconfinement by quarks and the Polyakov potential, we see a decoupling of the order parameters, which becomes stronger for large chemical potentials. We observe several different states of matter that can form, starting from a nucleon liquid which changes to a phase of chirally symmetric hadrons. Only at higher temperature these hadrons disappear and the quarks are the dominant degrees of freedom. Whether such a chirally symmetric hadronic phase can be the $N_c=3$ equivalent of the $N_c=\infty$ quarkyonic phase \cite{McLerran:2007qj} is still under extensive debate \cite{Lottini:2011zp,Bonanno:2011yr,Giacosa:2011uk}. In any case the high density part of the QCD phase diagram could have a rather rich phase structure to explore.\\

In the last part of this paper we discuss the properties of our models phase diagram at finite net-strange density. This aspect of QCD matter is not only interesting for heavy ion collisions and compact stars, but also for a comparison with lattice results extrapolated to finite $\mu_{B,S}$. We find that the location of the critical endpoint shifts to a slightly higher temperature for a finite net strangeness (lattice results).\\

We briefly discussed quantities that are sensitive on the chiral and/or deconfinement phase transition. In particular these are the strangeness baryon correlation factor $c_{BS}$ and the strangeness per baryon fraction $f_s$. Both show to be sensitive to the deconfined fraction on the system while $c_{BS}$ also shows a distinct peak at the chiral crossover at finite chemical potential. The advantage of extracting $f_s$ from the lattice is that it can be evaluated e.g. in a Taylor expansion, using only the first two non-zero expansion coefficients for the strange and light quark number susceptibilities.

\section*{Acknowledgments}
This work was supported by BMBF, HGS-hire and the Hessian LOEWE initiative through the Helmholtz International center for FAIR (HIC for FAIR). The authors thank P. Petreczky for fruitful discussions. The computational resources were provided by the LOEWE Frankfurt Center for Scientific Computing (LOEWE-CSC).
 
\bibliographystyle{apsrev4-1}

\end{document}